\documentclass[pra, showpacs, twocolumn, floatfix]{revtex4}
\usepackage{graphicx}
\usepackage{amsmath, amsfonts, amssymb, bm}
\begin{document}
\title{Cooling a quantum circuit via coupling to a multiqubit system}

\author{Mihai A. \surname{Macovei}}
\email{mihai.macovei@mpi-hd.mpg.de}

\affiliation{Max-Planck-Institut f\"{u}r Kernphysik, Saupfercheckweg
1, D-69117 Heidelberg, Germany}
\date{\today}
\begin{abstract}
The cooling effects of a quantum LC circuit coupled inductively with an 
ensemble of artificial qubits are investigated. The particles may decay 
independently or collectively through their interaction with the 
environmental vacuum electromagnetic field reservoir. For appropriate bath 
temperatures and the resonator's quality factors, we demonstrate an effective 
cooling well below the thermal background. In particular, we found that for 
larger samples the cooling efficiency is better for independent qubits. 
However, the cooling process can be faster for collectively interacting 
particles.
\end{abstract}
\maketitle
\section{Introduction}
The ability to cool interacting quantum systems below the values 
imposed by the thermal fluctuations of the environmental 
reservoir of each subsystem is actually of great interest \cite{rmp}.
For instance, a laser cooling technique for trapped particles 
exploiting quantum interference, or electromagnetically induced 
transparency, in a three-level atom was presented in \cite{chk}. 
There, by appropriately designing the absorption profile with a strong 
coupling laser, the cooling transitions induced by a cooling laser 
are enhanced while heating by resonant absorption is strongly 
suppressed. The experimental demonstration of ground state laser cooling 
with electromagnetically induced transparency was performed in 
\cite{eit_exp}. Recently, this idea was adopted to cool a nanomechanical 
resonator \cite{xe}. Further, the collective-emission-induced cooling of 
atoms in an optical cavity was also observed \cite{c_exp}.  Mechanical 
effects of light in optical resonators were studied in \cite{th1} while 
cavity-assisted nondestructive laser cooling of atomic qubits was 
analyzed in \cite{th2}. A laser cooling method that can be used at 
large detuning and low saturation to cool particles inside an optical 
cavity was proposed in Ref.~\cite{th3}. A significant speed-up of the 
cooling process was found in \cite{th4} while fast cooling of trapped
ions using the dynamical Stark shift was described in \cite{th5}.

Via engineering superconducting elements as artificial atoms and
coupling them to a photon field of a resonator or to vibrational
states of a nanomechanical resonator one can demonstrate 
interesting related phenomena such as single artificial atom lasing 
or cooling. In particular, schemes to ground-state cooling of
mechanical resonators were proposed in \cite{m_cool}. A flux qubit
was experimentally cooled \cite{fl_cool} using techniques somewhat
related to the well-known optical sideband cooling methods (see,
e.g., Ref.~\cite{rmp} and references therein). Lasing effects of a 
Josephson-junction charge qubit, embedded in a superconducting 
resonator, was experimentally demonstrated in \cite{l_exp}. 
Single-qubit lasing and cooling at the Rabi frequency was proposed 
in \cite{qu_cool}, while a mechanism of simultaneously cooling of 
an artificial atom and its neighboring quantum system was analyzed 
in \cite{sim_cool}. Few-qubit lasing in circuit QED was discussed in 
Ref.~\cite{l_col}. A LC oscillator can be cooled via its nonlinear 
coupling to a Josephson flux qubit \cite{mm_xt}. The cooling of a 
nanomechanical resonator via a Cooper pair box qubit has been recently 
suggested in Ref.~\cite{coop} while cooling carbon nanotubes to the 
phononic ground state with a constant electron current was achieved 
in \cite{cg}. Further interesting works on cooling micro- and 
nanomechanical resonators were presented in Refs.~
\cite{nm1,nm2,nm3,nm4,nm5,nm6,nm7,nm8,nm9,nm10}.
\begin{figure}[b]
\includegraphics[width=7cm]{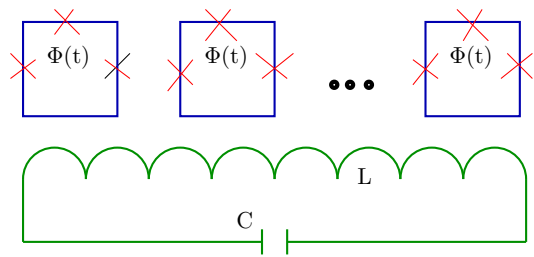}
\caption{\label{fig-1} An ensemble of $N$ independent three-junction flux 
qubits is coupled to a LC quantum circuit by their mutual inductances. An 
AC magnetic flux $\Phi(t)$ drives the qubits.}
\end{figure}

Here, we describe a cooling scheme via coupling a pumped multiparticle 
ensemble (i.e. artificial atoms or qubits) to a single mode of a quantum 
LC circuit (see Fig.~\ref{fig-1}). Our motivation is to present an 
efficient method allowing for a rapid cooling of the resonator mode. 
The multiqubit system can be formed by an independent $N$-particle sample 
or by collectively interacting $N$ particles. By independent, we mean 
that each particle spontaneously decays individually and all of them are 
maximally coupled with the oscillator mode and with the same phase. 
Collectively interacting particles means that their interactions are 
mediated by the environmental electromagnetic field reservoir such that 
their decay is of a collective nature. In this case, the particles are 
close to each other on a scale smaller than the emission wavelength and 
coupled with the same strength to the quantum oscillator mode. The 
advantages or disadvantages regarding the interparticle interactions to 
the cooling phenomena of the quantum oscillator degree of freedoms will 
be discussed in detail. In particular, we found that the cooling phenomenon 
is better for independently interacting qubits if the quantum dynamics of 
the LC oscillator is slower than that of the qubits. However, the cooling 
effects may occur faster for collectively interacting qubits. Apart from 
a fundamental interest, these systems have a great feature in various 
applications such as novel quantum sources of light (single photon sources, 
for instance), quantum processing of information or entanglement. However, 
at MHz frequency ranges thermal fluctuations affect considerably the LC 
oscillators, i.e., populate their energy levels and induce additional 
decoherences. Therefore, a suitable method to cool these systems can be very 
useful.

The paper is organized as follows. In Section II, we introduce the system 
of interest. Section III describes the obtained results. We finalize the 
article with conclusions presented in Section IV.

\section{Approach}
We describe the cooling effects of a quantum oscillator mode, i.e. 
a quantum LC circuit coupled inductively with a collection of 
two-level Josephson flux qubits (see Fig.~\ref{fig-1}). The two-level 
particles are pumped with a moderately intense magnetic flux and damped 
spontaneously via their interactions with the environmental electromagnetic 
field reservoir. The single-particle spontaneous decay rate is $\gamma$. 
Both subsystems interact with thermostats at effective temperatures $T_{1}$ 
and $T_{2}$. We shall consider that the particles are independent or 
collectively interacting. The frequency of the oscillator is much lower 
than the qubit's tunnel splitting, i.e. $\omega_{c} \ll \Delta$. Therefore, 
the qubit is driven with Rabi frequencies near resonance with the oscillator 
frequency that affect the oscillator, increasing its oscillation amplitude 
\cite{exp}. Near the symmetry point (i.e. the energy bias $\epsilon$ between 
the flux states is small) and after transformation to the qubit's eigenbasis, 
the Hamiltonian describing the multiqubit systems is:
\begin{eqnarray}
H&=&\sum^{N}_{i=1}\bigl\{\Delta E\sigma_{zi}/2 + 
\Omega_{0}\sin{2\theta}\sigma_{xi}\cos{(\omega t)}\bigr \} 
+ \omega_{c}a^{\dagger}a \nonumber \\
&-& g\sum^{N}_{i=1}\bigl(\cos{2\theta}\sigma_{zi}-
\sin{2\theta}\sigma_{xi}\bigr)(a + a^{\dagger}), \label{HMq}
\end{eqnarray}
where the first term describes the qubits, each with the transition 
frequency $\Delta E = \sqrt{\Delta^{2}+\epsilon^{2}}$, while the 
second one considers their driving by an applied AC magnetic flux with 
amplitude $\tilde \Omega_{0}=\Omega_{0}\sin{2\theta}$ and frequency 
$\omega$. Here, $\cot{2\theta}=\epsilon/\Delta$ with 
$\cos{2\theta}=\epsilon/\Delta E$ and $\sin{2\theta}=\Delta/\Delta E$. 
The last two terms describe the oscillator with frequency 
$\omega_{c}=1/\sqrt{LC}$ as well as the qubit-oscillator interaction, 
respectively. Here, $g \approx MI_{p}I_{c0}$, where $M$ is the mutual 
inductance, $I_{p}$ the magnitude of the persistent current in the qubit, 
and $I_{c0}=\sqrt{\omega_{c}/2L}$ the amplitude of the vacuum
fluctuations of the current in the LC oscillator. $a^{\dagger}$ and $a$
are the creation and annihilation operators corresponding to the
oscillator degrees of freedom, while $\sigma_{\alpha}$ $(\alpha \in
\{x,y,z\})$ are the Pauli matrices operating in the dressed flux
basis of the qubit subsystem. As $\Delta \gg \omega_{c}$, the 
transverse coupling in the Hamiltonian (\ref{HMq}) is transformed 
into a second-order longitudinal coupling by employing a 
Schrieffer-Wolff type transformation \cite{qu_cool,sal}, i.e. 
$U_{S}=\exp{(iS)}$ with 
\begin{eqnarray*}
S=(g/\Delta E)\sin{2\theta}(a+a^{\dagger})
\sum^{N}_{i=1}\sigma_{yi}.
\end{eqnarray*}
By further using the rotating wave approximation with respect to $\omega$ 
and diagonalizing the qubit term as well as applying the secular approximation, 
i.e. omitting terms oscillating with the generalized Rabi frequency, one arrives 
at the following Hamiltonian describing the interaction between the multiqubit 
system and the LC oscillator:
\begin{eqnarray}
H&=& \omega_{c}a^{\dagger}a + \sum^{N}_{i=1}\bigl\{\Omega R_{zi}/2 + 
\tilde g(R^{(i)}_{+-}a + a^{\dagger}R^{(i)}_{-+})  \nonumber \\
&+& g_{0}(aa^{\dagger} + a^{\dagger}a)R_{zi}/2 \bigr \}. \label{HM}
\end{eqnarray}
Here, we have further assumed that the generalized Rabi frequency is 
of the order of $\omega_{c}$, that is $\Omega \approx \omega_{c}$. 
In Eq.~(\ref{HM}), $\tilde g$=$g\cos{2\theta}\sin{2\xi}$ gives the 
qubit-oscillator coupling strength while 
$g_{0}$=$2g^{2}\cos{2\xi}\sin^{2}{2\theta}/\Delta E$ accounts for a 
small frequency shift of the qubit's frequency. Further,
\begin{subequations}
\label{sincos}
\begin{align}
&\cot{2\xi}=\delta \omega/\tilde \Omega_{0}, \\
&\cos^{2}{\xi}=[1+\delta\omega/\Omega]/2, \\
&\sin^{2}{\xi}=[1-\delta\omega/\Omega]/2.
\end{align}
\end{subequations}
where $\delta \omega =\Delta E -\omega$ and where 
$\Omega=\sqrt{(\delta \omega)^{2}+\tilde \Omega^{2}_{0}}$ stands for 
the generalized Rabi frequency. The dressed-state qubit operators 
$R^{(i)}_{\alpha\beta} = |\alpha\rangle_{i}{}_{i}\langle \beta|$
describe the internal transition in the $i$th particle between the 
dressed state $|\beta\rangle$ and $|\alpha\rangle$ for $\alpha \neq \beta$ 
and population for $\alpha=\beta$, $\{\alpha,\beta \in +,-\}$, and obey 
the standard commutation relations of su(2) algebra, that is
\begin{eqnarray*}
[R^{(j)}_{\alpha\beta },R^{(l)}_{\alpha^{\prime}\beta^{\prime }}] = 
\delta_{jl}\bigl(\delta _{\beta\alpha^{\prime}}R^{(j)}_{\alpha \beta ^{\prime }} - 
\delta _{\beta^{\prime}\alpha}R^{(j)}_{\alpha^{\prime }\beta}\bigr),
\end{eqnarray*}
where $\alpha,\beta \in \{+,-\}$. $R_{zi}=R^{(i)}_{++}-R^{(i)}_{--}$ 
is the dressed-state inversion operator for the $i$th particle.

In the mean-field, dipole, Born-Markov and secular approximations, the 
combined system is characterized by the following master equation:
\begin{eqnarray}
\frac{d}{dt}\rho + i[H,\rho]= - \Lambda_{a}\rho - \Lambda_{c}\rho. \label{ME} 
\end{eqnarray}
The quantum dissipation due to spontaneous emission into surrounding electromagnetic 
field reservoir is described by the $\Lambda_{a}\rho$ term which for $N$ independent 
qubits can be represented as follows:
\begin{eqnarray}
\Lambda_{a}\rho &=&\sum^{N}_{i=1}\bigl \{\gamma_{0}[R_{zi},R_{zi}\rho] + 
\gamma_{+}[R^{(i)}_{+-},R^{(i)}_{-+}\rho] \nonumber \\
&+& \gamma_{-}[R^{(i)}_{-+},R^{(i)}_{+-}\rho] \bigr \} + H.c.. \label{DI}
\end{eqnarray}
For $N$ nonindependent radiators, i.e. for collectively interacting 
particles, the corresponding damping is:
\begin{eqnarray}
\Lambda_{a}\rho &=&\sum^{N}_{i,j=1}\bigl \{\gamma_{0}[R_{zi},R_{zj}\rho] + 
\gamma_{+}[R^{(i)}_{+-},R^{(j)}_{-+}\rho] \nonumber \\
&+& \gamma_{-}[R^{(i)}_{-+},R^{(j)}_{+-}\rho] \bigr \} + H.c.. \label{DNI}
\end{eqnarray}
The damping rates are given by the following expressions:
\begin{subequations}
\label{decr}
\begin{align}
&\gamma_{+}=\gamma\cos^{4}{\xi}, \\
&\gamma_{-}=\gamma\sin^{4}{\xi}, \\
&\gamma_{0}=\gamma \sin^{2}{2\xi}/4.
\end{align}
\end{subequations}
The last term in Eq.~(\ref{ME}) characterizes the damping of the quantum 
oscillator mode and is given as follows:
\begin{eqnarray}
\Lambda_{c}\rho = \kappa\bigl(1+\bar n(\omega_{c})\bigr)[a^{\dagger},a\rho] + 
\kappa \bar n(\omega_{c})[a,a^{\dagger}\rho] + H.c.. \label{DC}
\end{eqnarray}
Here, $\bar n(\omega_{c})$ is the mean thermal photon number corresponding 
to the resonator frequency $\omega_{c}$ while $\kappa$ is the resonator 
decay rate. We have omitted the coherent part of the dipole-dipole interaction
in Eq.~(\ref{ME}), which is justified if the Rabi frequency dominates over the 
dipole-dipole induced energy shifts.

A general analytical solution of Eq.~(\ref{ME}) is not evident. 
However, one can obtain its solution for different regimes of interest, 
namely in the bad or good cavity limit. Therefore, in the next Section, 
we proceed by investigating the properties of Eq.~(\ref{ME}) when the 
qubit's quantum dynamics is faster than the one of the quantum oscillator, 
i.e. in the good cavity limit \cite{sz}.

\section{Results and discussions}
We assume a moderately intense pumping field, i.e., 
$\Omega \gg \{\gamma,\tilde g\sqrt{N}\}$ 
and a high quality resonator such that $\gamma \gg \tilde g\sqrt{N} \gg \kappa$ 
for $N$ independent particles or $\Omega \gg \{N\gamma,\tilde g\sqrt{N}\}$, and 
$N\gamma \gg \tilde g\sqrt{N} \gg \kappa$ for $N$ collectively interacting particles. 
Therefore, in this case, the qubit subsystem achieves its steady-state on a time 
scale faster than the resonator field and, thus, the qubit variables can be eliminated 
to arrive at a master equation for the resonator field mode alone: 
\begin{eqnarray}
\dot \rho = -\Gamma_{-}\{a^{\dagger}a\rho - a\rho a^{\dagger}\}
-\Gamma_{+}\{aa^{\dagger}\rho - a^{\dagger}\rho a\} + H.c., \label{FME} 
\end{eqnarray}
where an overdot means differentiation with respect to time and
\begin{subequations}
\label{cdr}
\begin{align}
&\Gamma_{-}=\kappa\bigl (1+ \bar n(\omega_{c})\bigr) + B, \\
&\Gamma_{+}=\kappa \bar n(\omega_{c}) + A.
\end{align}
\end{subequations}
The physical meaning of the parameters in Eq.~(\ref{FME}) is as
follows: $\Gamma_{+}~(\Gamma_{-})$ describes the process of increasing 
(decreasing) of the photon number in the resonator mode. The interplay 
between $\Gamma_{+}$ and $\Gamma_{-}$ leads to lasing or cooling of the 
quantum LC circuit.

For an independent $N$ particle system one has:
\begin{eqnarray}
A=\frac{\tilde g^{2}N}{\Gamma_{\perp}}\langle R_{++}\rangle,~ {\rm and}~
B=\frac{\tilde g^{2}N}{\Gamma_{\perp}}\langle R_{--}\rangle, \label{ABI}
\end{eqnarray}
where 
\begin{eqnarray}
\label{ddr}
\Gamma_{\perp}=4\gamma_{0}+\gamma_{+}+\gamma_{-}. 
\end{eqnarray}
We have considered here that $\langle R^{(i)}_{\alpha\beta}\rangle$ are identical 
for all $i \in \{1,2,\cdots,N\}$ and, thus, 
$\langle R^{(i)}_{\alpha\beta}\rangle \equiv \langle R_{\alpha\beta}\rangle$ where 
$\{\alpha,\beta \in +,-\}$. The expectation values for $\langle R_{\alpha\beta}\rangle$ 
are calculated in the absence of the resonator mode. Therefore, from Eq.~(\ref{DI}), 
we have:
\begin{eqnarray}
\langle R_{++}\rangle= \frac{\gamma_{-}}{\gamma_{-}+\gamma_{+}}~~ {\rm and}~~ 
\langle R_{--}\rangle= \frac{\gamma_{+}}{\gamma_{-}+\gamma_{+}}. 
\label{sacr}
\end{eqnarray}
The expressions (\ref{ABI}) are valid for any $N$ satisfying the restrictions 
imposed in the beginning of the Section including $N=1$.

For an ensemble of collectively interacting $N$ particles, we obtain 
the following relations for $A$ and $B$:
\begin{eqnarray}
\label{ABC}
A=\frac{\tilde g^{2}}{\tilde \Gamma_{\perp}}\langle R_{+-}R_{-+}\rangle,~{\rm and}~
B=\frac{\tilde g^{2}}{\tilde \Gamma_{\perp}}\langle R_{-+}R_{+-}\rangle, 
\end{eqnarray}
where the collective decay rate is given as follows:
\begin{eqnarray}
\label{cddr}
\tilde \Gamma_{\perp}= \Gamma_{\perp} + 
(\gamma_{-}-\gamma_{+})\langle R_{z}\rangle.
\end{eqnarray}
One can observe here that the decay rate $\tilde \Gamma_{\perp}$ has a 
contribution arising from all particles, i.e., the term proportional to 
$\langle R_{z}\rangle$. Here, in contrast to independent qubits, collective 
operators were introduced, that is, 
$R_{\alpha\beta} = \sum^{N}_{i=1}R^{(i)}_{\alpha\beta}$. Note 
that to obtain Eqs.~(\ref{ABC}), we decoupled the involved multiparticle 
correlators- an approximation valid for larger $N$, i.e., $N\gg 1$.
However, the corresponding expressions for $N=1$ are identical to 
Eqs.~(\ref{ABI}) but with the single-particle decay rate $\Gamma_{\perp}$ 
instead of collective ones. The steady-state expectation values for the 
collective correlators entering into the above expressions, i.e. Eqs.~(\ref{ABC}), 
can be estimated from the steady-state solution of the master equation 
Eq.~(\ref{DNI}) describing the strongly driven particles in the absence of the 
resonator \cite{bsq,xmek}:
\begin{subequations}
\label{sscr}
\begin{align}
&\langle R_{z}\rangle=\frac{N(1-f^{2+N})+f(N+2)(f^{N}-1)}{(f-1)(f^{N+1}-1)}, \\
&\langle R_{+-}R_{-+}\rangle=\frac{1}{1-f}\langle R_{z}\rangle, \\
&\langle R_{-+}R_{+-}\rangle=\frac{f}{1-f}\langle R_{z}\rangle, 
\end{align}
\end{subequations}
where $f=\gamma_{+}/\gamma_{-}$. For $N=1$ one obtains Eqs.~(\ref{sacr}).

The Eq.~(\ref{FME}) has an exact steady-state solution. For instance, the steady-state 
expectation values for the diagonal elements of Eq.~(\ref{FME}) can be obtained from 
the relation:
\begin{eqnarray}
\rho = Z^{-1}\exp{[-\alpha a^{\dagger}a]},
\label{SS}
\end{eqnarray}
where $\alpha=\ln(\eta)$ with $\eta = \Gamma_{-}/\Gamma_{+}$ and $Z$ 
is determined by the requirement $\rm Tr\{\rho\}=1$. Evidently, the 
expectation values of the operators needed for evaluating the properties 
of the quantum oscillator are obtained from Eq.~(\ref{SS}). In particular, 
the oscillator mean photon number, i.e. $\langle n\rangle \equiv 
\langle a^{\dagger}a\rangle = {\rm Tr}\{a^{\dagger}a\rho\}$, and its 
second-order correlations can be determined from the following expressions:
\begin{subequations}
\label{ssfcr}
\begin{align}
&\langle a^{\dagger}a\rangle = \frac{1}{\eta-1}, \\
&\langle a^{\dagger^{2}}a^{2}\rangle = \frac{2}{(\eta-1)^{2}},
\end{align}
\end{subequations}
to such an extent that the photon second-order correlation function, i.e. 
\begin{eqnarray*}
g^{(2)}(0)=\frac{\langle a^{\dagger^{2}}a^{2}\rangle}{\langle a^{\dagger}a\rangle^{2}},
\end{eqnarray*}
equals with $2$ which means that the photon statistics is always super-Poissonian.
Note that the mean photon numbers obtained with the help of Eq.~(\ref{FME}) or
Eq.~(\ref{SS}) should be below the photon saturation number $n_{0}$ which for 
$N$ interacting qubits reads approximately as:
\begin{eqnarray}
n_{0}=\Gamma_{\perp}(\gamma_{+}+\gamma_{-})/(\tilde g^{2}N).
\label{n0}
\end{eqnarray}
\begin{figure}[t]
\includegraphics[width=7cm]{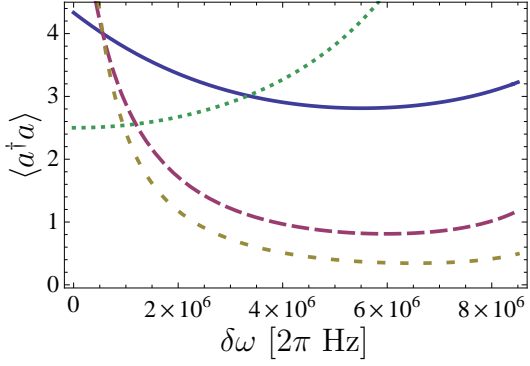}
\caption{\label{fig-2}(color online) The mean photon number $\langle n\rangle$ into 
the quantum circuit as function of $\delta \omega$ and different numbers of independent 
qubits. The solid line is for $N=1$, the long-dashed line stands for $N=10$, 
while the short-dashed curve corresponds to $N=30$. The dotted curve shows 
the saturation photon number $n_{0}$ for $N=30$ qubits. Here, $\bar n(\omega_{c})=4$, 
$\Delta/2\pi=3\cdot 10^{9}{\rm Hz}$, $\epsilon=0.01\Delta$, $g/2\pi=10^{6}{\rm Hz}$, 
$\omega_{c}/2\pi=10^{7}{\rm Hz}$, $\gamma/2\pi=10^{5}{\rm Hz}$, 
$\kappa/2\pi=10^{3}{\rm Hz}$ and 
$\tilde \Omega_{0}=\sqrt{\Omega^{2}-(\delta\omega)^{2}}$.}
\end{figure}

Fig.~(\ref{fig-2}) depicts the mean photon number in the oscillator mode 
which is coupled with $N$ independent qubits. We have used typical parameters 
here (see, for instance, \cite{exp}). To elucidate the role of many particles 
regarding the cooling issue, we fix the involved parameters and change the number 
of qubits. Already for $N=10$ particles, the cooling efficiency is significantly 
improved in comparison to the single-qubit case, i.e. $N=1$. Better cooling can 
be achieved, that is $\langle n\rangle \ll \bar n(\omega_{c})$, by increasing 
further the number of qubits (see the short-dashed curve in Fig.~\ref{fig-2}). 
Evidently, the qubits are in their lower dressed-state when cooling occurs, i.e. 
$\langle R_{--}\rangle > \langle R_{++}\rangle$. The diagram showing the energy 
levels of the qubit and oscillator indicating the cooling cycle with photon 
emission/absorption can be found in Refs.~\cite{nm6,nm7}.

\begin{figure}[t]
\includegraphics[width=7cm]{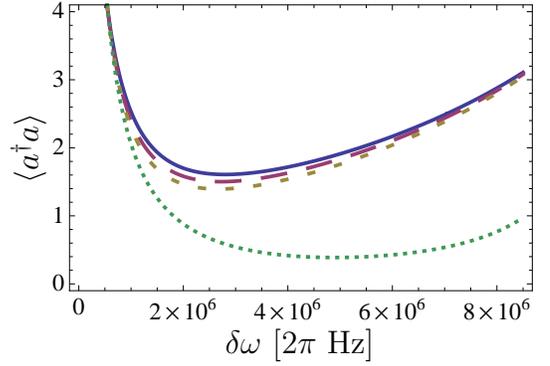}
\caption{\label{fig-3}(color online) The mean photon number $\langle a^{\dagger}a\rangle$  
of the quantum oscillator as the function of $\delta \omega$ and different numbers of 
collectively interacting qubits. The solid line is for $N=10$ while the long-dashed line 
stands for $N=15$. The short-dashed curve corresponds to $N=30$ while the dotted one 
to $N=30$ but $\kappa/2\pi=10^{2}{\rm Hz}$. Other parameters are the 
same as in Fig.~({\ref{fig-2}}).}
\end{figure}

Further, we turn to cooling effects via collectively interacting particles.
Fig.~(\ref{fig-3}) shows the mean photon number in the quantum oscillator 
mode as the function of $\delta \omega$. The mean photon number $\langle n\rangle$ 
is well below the thermal mean photon number $\bar n(\omega_{c})$, however the 
cooling is not so significant as for independent qubits (compare the short-dashed 
curves in Fig.~\ref{fig-2} and Fig.~\ref{fig-3}). The reason is that the decay 
rate $\tilde \Gamma_{\perp}$ is dependent on the number of qubits and, thus, 
$\langle n\rangle = \Gamma_{+}/(\Gamma_{-}-\Gamma_{+})$ is smaller than the 
corresponding one for independent qubits since $\Gamma_{-}-\Gamma_{+}=\kappa -
\tilde g^{2}\langle R_{z}\rangle/\tilde \Gamma_{\perp}$ where 
$\langle R_{z}\rangle$ is given by Eq.~(\ref{sscr}) (in other words, for 
collectively interacting particles, we do not have a factor $N$ in the 
denominator). However, adjusting the involved parameters, one can improve 
the cooling efficiency in general (see the dotted line in Fig.~\ref{fig-3}). 
Note that the coupling of qubits can be controlled \cite{contr}.

Finally, we discuss the time scaling for the cooling phenomenon. We observe 
that cooling rates depend on the number of qubits and, therefore, the 
cooling may occur faster in both schemes. However, the faster decay rate 
in our approach is the qubit spontaneous emission. Thus, the cooling 
phenomena can not occur faster than $\gamma^{-1}$ for independent qubits 
or $(N\gamma)^{-1}$ for collectively interacting particles, respectively. 
Therefore, in general, the cooling processes are faster for collectively 
interacting particles.

\section{Summary}
In summary, we described a scheme that is able to cool a quantum 
LC circuit coupled inductively to externally pumped artificial 
particles (Josephson flux qubits) and damped through their 
interaction with the environmental electromagnetic field reservoir. 
The qubits may interact collectively or they are independent. 
If the qubit’s dynamics is faster than the one of the LC oscillator, the 
cooling of the oscillator’s degrees of freedom occurs when controlling
the qubit quantum dynamics. We found that the cooling phenomenon is 
better for an ensemble of independent qubits. However, in general, the 
cooling processes are faster for collectively interacting particles.

\end{document}